\documentclass[10pt, oneside]{article}   	
\usepackage{geometry}                 
\usepackage{graphicx,multicol}
\usepackage[english]{babel}
\usepackage{amssymb,bm,hyperref,color}
\usepackage{cite}
\usepackage{multicol,booktabs}
\usepackage{nicefrac}
\usepackage{bbold}
\usepackage{comment}
\usepackage[english]{babel}
\usepackage[utf8]{inputenc}

\title{\sc\Large From Hole Theory to Quantum Field Theory:\\ \large Relativistic Fermions and the Role of Ettore Majorana (1933-1937)}
\author{Francesco Vissani\\ \small INFN, Laboratori Nazionali del Gran Sasso}
\date{}						

\begin{document}
\maketitle

\renewcommand{\abstractname}{\vskip-3mm Abstract}
\begin{abstract}

Between 1933 and 1937, the treatment of relativistic spin-$\nicefrac{1}{2}$ particles, initially rooted in Hole theory, evolved into the modern framework of quantum field theory. This paper reconstructs the crucial stages of that transition by examining the formal and physical progress of the numerous authors who shaped the field's modern formalism. This historical study traces the development of fermionic field theory in full, beginning with the foundational work of the 1920s, focussing on the results of the 1930s, and concluding with the influential synthesis of Wolfgang Pauli in 1941, the content of which has shaped the subsequent tradition.
Within this framework, particular emphasis is given to Ettore Majorana's 1937 quantisation procedure and argument for anti-commuting fermionic quantum fields. This study demonstrates that Majorana's work was not merely a technical variant, but the definitive rejection of the concept of negative energy solutions, whose conceptual clarity and educational value remain vital today.

\end{abstract}



\centerline{\footnotesize \sf Keywords: \sf history of quantum field theory, hole theory, second quantisation,  symmetric theory,  Majorana fermions}

\parskip0.4ex


\section{Introduction}\label{s:1}

The concept of the ``fermionic quantum field"—the modern operator formalism merging quantum mechanics with special relativity to describe matter—is a cornerstone of contemporary physics. While it fundamental characteristics, such as half-integer spin, Fermi-Dirac statistics, and the necessity of antiparticles, are now standard, the concept itself is one of the most intellectually demanding in theoretical high-energy physics. Its history reflects this inherent difficulty; far from a linear progression, the ``fermionic quantum field" emerged through a series of non-obvious conceptual leaps and formal struggles that remain among the least immediate to grasp even for the modern intellect.

A decisive phase in the construction of this intellectual tool occurred between 1933 and 1937, a period marked by the final conceptual and procedural advances required for its modern definition. While the era prior to 1933 has been explored in detail by Darrigol \cite{d86}, focusing on the heuristics of the early protagonists, and the period following 1937 has been addressed by Blum \cite{b14} regarding the definitive systematisation of the exclusion principle, the ``middle period" remains less explored. This paper addresses that gap by examining the critical five-year window culminating in 1937.
Specifically, it highlights the significance of Ettore Majorana’s final paper \cite{m37}, arguing that it has been a crucial, yet 
overlooked, pivot from early heuristic models to the modern understanding of fermionic quantum fields.
Table~\ref{tub1} provides a comprehensive chronology of major contributions to the modern fermionic quantum field concept during this era, which are the subject of detailed discussion in the subsequent sections of this paper.



To understand the delayed development of the fermion field concept, one must consider the specific priorities of the 1930s. At the time, theoretical efforts were largely consumed by the immediate challenges of relativistic electromagnetic theory and the newfound discovery of antiparticles in 1932. The emergence of infinite quantities---the ``divergences''---created a sense of crisis that prioritised investigating electron mass 
\cite{o30,w30,w39} and vacuum polarization \cite{h34,e36}
over the abstract ontology of fields. In this climate of urgent problem-solving, the deeper conceptual clarification of the fermion field was often sidelined. It is within this context of ``crisis'' that Majorana's work stands out, not merely as a formal variation, but as a 
deliberate rejection of Dirac's concept of negative energy solutions, that at the time was still very influential.

We begin in Sect.~\ref{s:2} by summarising the early era dominated by Dirac. Sect.~\ref{s:3} tracks the subsequent symmetric quantisation schemes, culminating with Heisenberg in 1934. Sect.~\ref{s:4}---the core of this work---illustrates the modifications introduced by Weisskopf, Pauli, and Majorana, reconstructing the latter's argument for anti-commuting fermion fields in modern terms. Finally, Sect.~\ref{s:5} traces the events leading to Pauli's 1941 influential synthesis, followed by a summary and final considerations in Sect.~\ref{s:6}.

\begin{table}[t]
\centering
\caption{\small Major contributions to the definition and clarification of the modern fermionic quantum field (1924--1941). The groupings correspond to the discussions in Sections \ref{s:2}, \ref{s:3}, \ref{s:4}, \ref{s:5}  of this paper.}\label{tub1}
\vskip3mm\small
\label{tab:contributions}
\begin{tabular}{@{}lll@{}}
\toprule
\textbf{Author} & \textbf{Major contribution(s)} & \textbf{Date} \\ \midrule
de Broglie & Matter-wave hypothesis & 1924 \\
Pauli & Exclusion principle & 1925 \\
Fermi; Dirac & Fermion statistics & 1926 \\
Jordan & Anti-commuting operator formalism & 1928 \\
Dirac & Relativistic wave equation; $\gamma$ matrices & 1928 \\
\textit{Dirac} & \textit{Physical origin of spin 1/2} & \\ 
Dirac & Hole theory & 1931 \\
Fermi & Description of matter particle creation ($\beta$-decay) & 1933 \\ \midrule
Fock & Construction of positive definite Hamiltonian & 1933 \\
Furry-Oppenheimer & Symmetric operatorial formalism & 1933 \\ 
Heisenberg & Symmetric quantisation & 1934 \\ \midrule
Pauli-Weisskopf & Canonical quantisation of scalar fields & 1934 \\
\textit{Pauli-Weisskopf} & \textit{Antiparticles without hole theory} & \\
Majorana & Overcoming the hole theory & 1937 \\
\textit{Majorana} & \textit{Argument for fermion statistics} & \\
\textit{Majorana} & \textit{New hypothesis on neutral particles} & \\
Ivanenko-Sokolov & Dirac-Heisenberg matches to Pauli-Weisskopf & 1937 \\
Racah & Remarks on neutral particles; discrete symmetries & 1937 \\
Kramers & $C$-matrix for general representation of $\gamma$ matrices & 1937 \\ \midrule
Fierz & Argument for spin-statistics connection & 1939 \\
de Wet & Argument for spin-statistics connection & 1940 \\
Pauli & General arguments on spin-statistics for any spin & 1940 \\
Pauli & Synthesis of fermionic quantum field theory & 1941 \\ \bottomrule
\end{tabular}
\end{table}

The historical scholarship on Ettore Majorana is extensive, with seminal works by Esposito~\cite{bs,bs1} and comprehensive Italian biographies by Bonolis 
\cite{bon1} and Guerra and Robotti \cite{guro,guro2} providing deep insight into his life and academic career. Due to his enduring mystique, narratives in his homeland have occasionally leaned toward a localized, person-centric focus. This paper seeks to broaden that perspective. By conducting a detailed technical analysis of Majorana’s final 1937 paper \cite{m37} alongside contemporaneous primary literature, we contextualise his work within the  landscape of 1930s theoretical physics. Our aim is to reconstruct Majorana’s specific role in the collective evolution of field quantisation. To assist the modern reader in navigating the original text’s complexities, we have provided supplementary notes in Appendix~\ref{appa}.

We conclude this introduction 
with a note on the historical context.
The international scientific discourse during this period of growing nationalism involved significant linguistic and political considerations. German was initially predominant, but English began to emerge as the new standard, a process consolidated after the war. Leading theorists were often polyglots, and conferences remained open forums, enabling the exchange of ideas that this paper analyses. Review journals, such as the {\em Review of Modern Physics}, also gained prominence~\cite{lalli}, publishing key synthesis works by figures like Fermi \cite{f32}  
and Pauli \cite{p41}.\footnote{Note that, like many others of their generation, both authors moved to the United States;
Fermi before the war (1938-39) while Pauli,
who had served as a visiting professor at Michigan and Princeton during the 1930s, 
moved just after it began (1940). 
A fascinating account of this process comes from Laura Fermi \cite{lf}.}  Comments on the specific internal situation of physics in Italy during the time of Majorana's last work are collected in Appendix~\ref{appb}.

\section{From quantum matter waves  to the beginning of Dirac's era (1924-1933)}\label{s:2}

This section provides an overview of the early stages in the history of quantised matter fields, 
which spans from 1924 to 1933. To refer to this early period we borrow  the term `quantised matter waves' from \cite{d86}, 
as opposed to ‘fermionic quantum fields’, the conceptual destination of this theoretical evolution.
I will summarise some significant facts without going into detail, unlike in the rest of the paper; for an exhaustive
exhaustive discussion of this period, see \cite{d86} and \cite{b14}.  
After presenting some key figures and their major contribution (sect.~\ref{2:1}), I highlight the innovations brought about by the context of relativity, which was introduced and then shaped in a peculiar way by Paul Dirac (sect.~\ref{2:2}). Fermi's theory of beta decay is based on the ideas of Dirac and 
the formalism of Jordan {\em et al}. 
His expression for the relativistic fermionic quantum field can be considered both a crowning achievement and the completion of the first period (sect.~\ref{2:3}).

\subsection{The pioneers: de Broglie, Pauli, Jordan and Dirac}\label{2:1}

Following \cite{d86}, we begin recalling 
the propulsive roles of de Broglie \cite{luigi}, for the description of the state of the electron in terms of  matter wave, and of Jordan, for the operatorial formalism that allowed an effective treatment of discrete particle states \cite{jk,jw}. 
For the current discussion of matter particles, I deem that also Pauli should be recognised as having played a similar role, if not else, for his concept of 
\begin{quote}
\em\small ``two-valuedness not describable classically''  
\end{quote}
namely, an intrinsic dual nature of the 
electron,  which is expressed by a  matter wave with two components. 
In practice, to explain all atomic spectra, it was necessary to go beyond the limits of classical analogies and  
use something fundamentally different -- an entirely quantistic feature.
This, associated with the formulation of the exclusion principle \cite{p25} for multi-electron states, 
paved the way for many subsequent and significant developments, including those of Jordan.

In this connection, it is interesting to comment on Dirac's role.  
On the one hand, he was the first to devise a quantum theory of radiation \cite{d27}, and also the one who formulated the relativistic wave equation of the electron \cite{d28},  which contributed to  put him at the forefront of the discussion on quantum field theory for many years. 
On the other hand, Dirac preferred to reason about the symmetry properties of systems of identical particles rather than adopt the aforementioned powerful operator formalism to treat electrons: this ultimately led to subsequent progresses being made by other scientists. Ref.~\cite{d86} advanced a convincing conjecture: in his research, Dirac aspired to explanations and formalisms\footnote{This attitude manifests itself, e.g., in \cite{d27} where he states that the number $N$ of photons and their phase $\theta$ are canonically conjugated and must obey the commutation rule $[{N}, \theta]=i\hbar$, based on an analogy with momentum and position. This heuristics gains a formula to proceed in the discussion, but considering the purely formal solution $\theta=-i\hbar d/d{N}$ it is evident that it cannot be given mathematical meaning.}  
 that, in one way or another, corresponded to the classical case, but  --  as history has shown  --   it was precisely the formalism developed by Jordan and others that made it possible to describe the substantial departure from the classical case, that allowed the construction of the quantum theory of the electron.

Let us now proceed to describe a crucial theoretical step by him which will guide the entire field until 1937: the
 {\em interpretation} of Dirac equation.

\subsection{Dirac and the relativistic electrons. Hole theory}\label{2:2}
\paragraph{Relativistic wave equation and the spin $\nicefrac{1}{2}$}
The original motivation for the relativistic quantum theory of the electron (Dirac's wave equation) \cite{d28} was as follows:
\begin{quote}
\small\em ``It appears that the simplest Hamiltonian for a point-charge electron satisfying the requirements of both relativity and the general transformation theory leads to an explanation of all duplexity phenomena without further assumption.''
\end{quote}
Thus, the main original point of the work titled 
{\em The Quantum Theory of the Electron}
was to elaborate the point of Pauli,  that is, to explain the 
existence of the spin.
We recall the explicit expression here, as it will be useful in later discussions:
\begin{equation}\label{dira}
i\hbar \frac{d\psi}{dt} =  \hat{H}\psi
\qquad \mbox{ where }\hat{H}=  \vec{\alpha} c\, (-i\hbar \vec{\nabla })  +  \beta \, m c^2
\end{equation}
where we have adopted the notation of \S21 of \cite{lando} and explicitly indicated the constants $\hbar$ and $c$. Of course, $\psi$ is the wave function of the 
electron and  $m$ identifies the mass of the electron. The $4 \times 4$ matrices  $\vec{\alpha}$ and $\beta$ are  Hermitian and anticommuting among them; they can be chosen in several ways, called `representations', and  correspond to the well-known gamma matrices according to $\gamma^0=\beta$  and $\vec{\gamma}=\beta \vec{\alpha}$. 
E.g., with the representation adopted by Majorana in \cite{m37} 
the $\vec{\alpha}$s are real, $\beta$  is imaginary, 
therefore the 
$\gamma^\mu$ with $\mu=0,1,2,3$ are all imaginary: 
thus Dirac equation is equivalent to a real differential equation.
 For an introduction, see Appendix~\ref{appa}.

\paragraph{A key distinction}
In order to appreciate the following discussion, it is very important to make clear from the outset the distinction between {\em wave quantum theory} —based on numerical quantities—and {\em field theory}—based instead on operators. A convenient language to define the difference 
was proposed by Dirac in 1926 \cite{d26}, from which we draw the following passage
\begin{quote}
\small \em ``The fact that the variables used for describing a dynamical system do not
satisfy the commutative law means, of course, that they are not numbers in
the sense of the word previously used in mathematics. To distinguish the two
kinds of numbers, we shall call the quantum variables q-numbers and the
numbers of classical mathematics which satisfy the commutative law c-numbers,
while the word number alone will be used to denote either a q-number or a
c-number.''
\end{quote}
We take this opportunity to recall that the Hamiltonian of eq.~\ref{dira} was introduced immediately after Dirac's equation by Tetrode \cite{ht28}. There is no doubt that he uses wave theory -- i.e. c-numbers -- even if he does not adopt the technical language, as is evident from the following passage, which I report in translation
\begin{quote}
\small \em ``As in classical theory, they should always be ``ordinary” quantities, i.e. they should differ from the unit matrix only by a factor and are therefore commutable with all matrices from the group of \(\gamma _{\mu }\) during multiplication.''
\end{quote}



\paragraph{The negative energy solutions}
The other outstanding feature of this wave equation is the presence of seemingly spurious solutions that mirror the expected ones in the region of negative energies. In a classical or non-relativistic framework, these are absent; however, 
their existence in the new context—that of relativistic wave theory—implied an alarming absence of a ground state, as electrons would be expected to transition indefinitely toward lower and lower energy states, making matter inherently unstable.
Dirac devoted considerable thought to this issue and then proposed that, based on the Pauli exclusion principle, all negative-energy electron states are occupied. This hypothetical ground state 
is known 
as the `Dirac sea' hypothesis -- an evocative term that conveniently summarises an interesting aspect of 
Dirac's heuristics or ontology.\footnote{See \cite{ponte} for a broader discussion. We will use the retronym `Dirac sea' wherever it helps to clarify matters; according to resource 
 {\tt ngram} \cite{ngram}, this became popular after WWII.}
  Interestingly, Dirac did not use this term himself, but instead used the expression `Hole theory', which describes an observable consequence of his conceptual setup; this notion and terminology will be used by leading physicists for nearly a decade.\footnote{To understand its enduring influence, one example suffices: as recently as 1950, Foldy and Wouthuysen's renowned work \cite{fw} discusses negative energies. Dirac's celebrated book \cite{paulo}, which is still widely used in universities, introduces the concepts of `hole' in \S~65 and of `Dirac sea' in \S~73, although the latter term is not employed.}

In his first attempt, Dirac proposes  to identify the holes  in the Dirac sea with protons \cite{d30}. In this  way he
 aimed at explaining all matter particles then known with a single theory seconding a monistic philosophical attitude
 (as it is well known, before the discovery of the neutron the only  and widely  accepted 
 theory of the composition of atomic nuclei was that they are made of electrons and protons).
Oppenheimer \cite{o301}, Tamm \cite{t30}  
and  Weyl  \cite{w31} participated  in this stage of the discussion.
The first two pointed out that electron-proton annihilation would be too fast; 
the last one entertained the hope  that this annihilation process could have
 explained the workings of stars, however, he also pointed out that the theory would have to be profoundly modified, since it predicted that electrons and protons had the same mass. This helped Dirac and according to \cite{d86} it prepared the evolutions towards `symmetric quantisation'. 

\paragraph{Hole theory and Dirac sea}
Dirac's next proposal, 
 to conceive the `holes' as  new and still unobserved particles  \cite{d31} with the same mass of the electrons and opposite charge,  
 anticipates Anderson's discovery  of the positron \cite{a32} by one year.
This successful prediction 
and the recognition in 1933 of the Nobel committee accreted greatly his scientific authority, and it would not an exaggeration to  claim that  this series of events marked the beginning of a sort of 
a `Dirac era'; compare with chap.~15 of~\cite{hund}. One important remarks is in order: 
the term ``anti-electron'' identifies the theoretical construct described just above, i.e., the hole in Dirac sea
and the title of Dirac's Nobel lecture {\em The Quantum Theory of the Electrons and Positrons}
\cite{d33} suggests the identification of his theoretical construct
with the particle observed by Anderson, the ``positron''.

The search for alternative electron wave equations  
include Weyl's theory of the massless electron \cite{w29}, van der Warden systematic analysis \cite{vw29}, 
and  a proposal by Majorana\footnote{Here is a intriguing remark from a late 1960s paper \cite{d67}, that may help us to appreciate the originality and forward-thinking nature of Majorana. The first line of the paper informs us that  {\em ``Wave equations of arbitrary spin were first$^\ddagger$ put forward by Dirac (1936),''}, whilst the footnote text reads as follows: {\em `` $^\ddagger$First, that is, except for the pioneer work of Majorana (1932) whose paper has passed into obscurity.''}
It is worth noting that in 1932 Majorana sought to eliminate negative energy solutions within the framework of relativistic wave mechanics; he would achieve this goal definitively only in 1937 by moving to the framework of quantum field theory.}
\cite{m32} whose (infinite dimensional) equation obeys the principles of relativity, but lacks 
 negative energy solutions. While this study has no applications in physics, it is of great mathematical interest. 
We would like to point out that, although Majorana based his research into this matter solely on general principles, it was extremely deep, as revealed by an analysis of his notebooks \cite{e09,e12}.

\subsection{Fermi and  ``second quantisation''}\label{2:3}
At the end of 1933 the Italian journal {\sf La Ricerca Scientifica}, published by the National Research Council (CNR), featured 
the first quantitative theory of beta decay~\cite{rs33} later re-published in an extended form~\cite{nc34}.
Arguably, the most impressive feature of this article is a revolutionary application of the formalism of Jordan and collaborator: the description of  the creation of new matter particles.\footnote{Darrigol \cite{d86} quotes an interesting  description of Jordan of this application: {\em ``In Democritus' representation
each individual atom had a definite destiny and possessed in its indestructibility and invariability the permanent guarantee of its lasting identity; 
while the electrons and other elementary particles of the modern physicist, aside from their destruction and conversion properties, possess no individuality''.} 
This is quite impressive, but to clearly discern historical transitions, avoiding the effects of hindsight, it us useful to note that these statements appeared eleven years after Fermi's work first demonstrated how to model the creation of matter particles.
In other words, I am not convinced that the great potential of Jordan's formalism should automatically imply the entire credit for the subsequent formal or  conceptual developments. For a similar opinion on this matter, see~\cite{yang}.}
Thanks to this work, the previous ideas of Ambarzumian, 
Ivanenko and Francis Perrin \cite{ai}, \cite{cr} -- inspired by de Broglie's vision  -- 
receive a concrete mathematical shape. 
See \cite{yang} for a discussion  of Fermi's paper aimed at modern readers.  

For what concerns the present discussion of fermionic quantum fields, and as emphasised in \cite{fermi,bang}, 
it should be noted that Fermi uses the following type of quantum electron field
\begin{equation}\label{cicciolo}
\bm\psi=\sum_s \bm a_{s}\ \psi_s
\end{equation}
Here, $\bm a_s$ are annihilation operators\footnote{These operators describe the annihilation of a spin 1/2 particle (an electron), which corresponds to a normalised wavefunction. Their Hermitian conjugates, on the other hand, describe the creation of a particle. In the bra-ket notation, this is expressed as the following condition on their non-zero matrix elements:
$\langle 0| \bm a_s | 1_s\rangle=\langle 1_s| \bm a_s^\dagger | 0 \rangle= 1$, where $|0\rangle$ is the vector of the vacuum state and $|1_s\rangle$ is the vector of the single particle state. Thus, $\psi$ and $\bm\psi$ have the same dimensions, namely, the inverse of a 
square root of a volume.} of Jordan, Klein, Wigner, Fock, Heisenberg,
and the sum is extended to all positive {\em and} negative energy states~$s$.
An important remark on the notation adopted throughout this paper: 
\begin{quote}
\sc boldface characters distinguish quantised fields from matter waves
\end{quote} 
For instance, in eq.~\ref{cicciolo},  $\bm\psi$ is the field (the q-number)
while $\psi$ is the wave (the c-number).
Throughout this paper, I adopt this convention that differs from common practice at the time, when the distinction was usually determined by context; I hope this approach will contribute to conceptual clarity. 
Incidentally, the notation   $\bm a^*$  was commonly used at the time to represent the Hermitian conjugate of the operator  $\bm a$. 
By using a boldface symbol, we avoid any confusion with the symbol for complex conjugation.

Both $\bm\psi$ and $\psi$
satisfy Dirac's equation for the given particle  --  in this case, the electron.
The formalism that describes the creation and annihilation of particles presents a close correspondence between fields and wave functions, which is the somewhat ``spontaneous position'' as formalised in eq.~\ref{cicciolo}.
It is essential to note that this construct works in tandem with the  
Dirac sea hypothesis as explicitly declared -- albeit with little emphasis -- in \cite{nc34}: see again \cite{fermi,bang} for a detailed discussion of this point.  Thus, Fermi’s work represents a sophisticated operational upgrade of Hole theory rather than its conceptual replacement.
Therefore, Fermi's article is, among the other things, also a conclusive development of the stage of quantum matter fields:  
Two concluding notes:
\begin{itemize}
\item In 1933, Fermi speaks of the application of the {\em Dirac-Jordan-Klein method of ``quantised probability amplitudes''} \cite{rs33}.
In the subsequent works \cite{nc34}, he cites \cite{jk} and \cite{h31} and speaks instead of:
{\em Dirac-Jordan-Klein method, called of ``second quantisation''.}
The last denomination 
seems preferable, and it is still in use in nuclear physics and other non-relativistic contexts.
\item  Fermi's theory will give rise to many valuable applications and developments. A neat example (the first) is the paper of 
one of his collaborators, Wick \cite{wick}, that proceeds in the discussion showing that the theory describes the existence of $\beta^+$ (positron) emission, already observed but yet to be understood, and predicts the possibility of 
electron capture, which would soon be observed. 
(See \cite{bang} for further discussion.)
\end{itemize}
In summary, during this initial period, valuable ideas and insights were gathered that would find application over time. However, the first conceptual framework to  quantise relativistic fermions, which was sufficiently developed and that guided all subsequent discussions until it was conceptually superseded, was the Hole theory, i.e., the scheme based on the hypothesis of the existence of the Dirac sea and that heavily relied on the exclusion principle.

%


\section{Formal progresses. Symmetric quantisation (1933-1934)}\label{s:3}

This section marks a pivotal transition in the conceptual history of matter waves. We shall examine three foundational papers published between late 1933 and 1934 that proposed critical modifications to Dirac’s hole theory. While these works emerged within a narrow six-month window, they collectively established a framework for what is now retrospectively termed ``symmetric quantisation." 
An essential result of the discussion,  that will lead to a significant progress toward fermionic quantum field, 
can be summarised in the form of the following implication
\begin{center}
\sc\small The free Hamiltonian  built in terms of  quantum fields can be made positive-definite 
modifying the definition in eq.~\ref{cicciolo}, in a way that justifies the term ``symmetric quantisation''
\end{center}
It should be noted, however, that the actors themselves did not use this label at the time but rather ``Dirac–Heisenberg quantisation"; however, the term ``symmetric quantisation" serves as a useful analytical category for the kind of procedural consistency they shared.

In what follows, I have chosen to adhere to the original notations employed by the authors. While this philological attitude requires navigating shifting symbols, it is intended to demonstrate that the formalism was not ``for free"; the evolution of notation—culminating in Pauli’s 1935 pedagogical synthesis—was inextricably linked to the struggle to interpret the physical reality of the electron-positron field.

Despite the importance of these results, 
their full implications were not immediately embraced. None of these papers explicitly emphasized what has since become a standard conceptual point: that the consistent application of this symmetric procedure allows one to bypass the physical hypothesis of negative-energy states entirely. As evidenced by a testimony from Pauli, these developments were viewed at the time less as a revolutionary departure and more as a formal refinement or ``variant" of Dirac’s original hole theory. However,  the results of this discussion anticipated modern positions  --  or formally coincided with them, as in the case of Heisenberg's formulation.

\subsection{Fock 1933  \cite{f33}}

Fock, in his paper of November 21, 1933    \cite{f33} remarks that, in the relativistic case that characterises the    Hamiltonian density that corresponds to Dirac equation,  there are reasons to modify the form of the quantised matter field of the electron. 
%
%
He proceeds rather formally, noting that 
the Dirac quantised field, written following the prescriptions of Jordan and Wigner, includes the part with positive energies $\bm\psi_+$ and the part with negative energies $\bm\psi_-$.
As Schr\"odinger noted, each observable constructed with them
includes a  `even' component, without transitions between the two parts, and a  `odd'  component, with transitions.
After observing that Dirac's free Hamiltonian is even, 
and inspired by a paper of Heisenberg \cite{h31} on the symmetry between empty and filled states in 
atomic shells, 
Fock proposes to reformulate the theory with the new field
%
\begin{equation}\bm\varphi=\bm\psi_+ + (\bm\psi_-)^*\end{equation} 
and in doing so he finds that the energy operator assumes {\em only positive eigenvalues.} As is evident to a modern reader, some version of this idea will have a great future.

However, the specific position of Fock led to difficulties, such as the lack of gauge invariance which was promptly noted by Fock\footnote{It is worth recalling that it was V.~Fock the first who introduced the concept of gauge invariance of electromagnetic interactions as 
 local transformation U(1) of the electron field, as clarified in \cite{jo}; thus, his concern was not a minor one.} 
and underlined by various readers. The archive \cite{sith} reports  two reviews of the aforementioned paper of Fock:  
K.~Schr\"{o}der states: 
{\em ``This note represents an attempt to formulate Dirac's positron theory mathematically. The requirement of gauge invariance leads to considerable difficulties. Furthermore, disturbing infinities arise in the wave equations of electrons and positrons in configuration space.''}\  
O.~Klein (author of the famous calculation with Nishina and close collaborator of Jordan)  instead remarks:
{\em ``Based on the quantum mechanics of a `hole'~in a complete electron group developed by Heisenberg, a formulation for Dirac's hole theory is presented, in which electrons and positrons appear only in positive energy states. Certain difficulties, including the theory's lack of gauge invariance, are discussed.''}\ 
It is useful to note that the form of the field shown in the text,
in the general case of the spinor representation, does not even respect Lorentz invariance as the representation is not real; this point will be clarified thanks to Majorana's work.\footnote{As recalled in Appendix~\ref{appa},  in Majorana representation of the $\gamma$ matrices, the  Lorentz transformation
matrices acting on the spinors $\Lambda(\omega)$  are real.}
 
 With the paper of Fock, the formal devising of a new form of the quantised field was usefully initiated, but not yet perfected; the discussion of its meaning had not yet begun. In this 
specific assessment, I   
somewhat depart from Darrigol \cite{d86}, who instead argued that:  
\begin{quote}
{\em\small ``Apparently, only Fock and his Russian
collaborators realized at that time that this was not only a formal
rewriting of Dirac's concepts, but a way to dry up the infinite electron
sea altogether''} 
\end{quote}
I have not found any statements attesting this level of awareness in the literature prior to Majorana's 1937 work.

\subsection{Furry-Oppenheimer 1933 \cite{fo33}}

Just 10 days after Fock's work, a   fully independent investigation of Furry and Oppenheimer appeared.
In section 2 of their work, a formal change is proposed, that of 
 \begin{quote}
\small \em
``regarding the emptiness and not the fullness of a state of negative kinetic energy as equivalent to the presence of a particle''
 \end{quote}
  In this way the  Hamiltonian built following Dirac, Jordan  and Wigner 
  has only positive terms,   up to an additive constant, formally infinite but without physical importance,
which could be regarded as the unobservable  
energy stored in the Dirac sea.
In this work the emphasis is on the identification of negative energy states with oppositely charged particles; it is stated that the 
 number of negative kinetic energies  states $\bm M$  can be interpreted 
  as $\bm{1-N}$.
  
A major aim of Oppenheimer and Furry is to suitably define bilinear operators of interest such as the number operator, the charge and the 
Hamiltonian density for free particles.
The observables are expressed in terms of the creation and destruction operators, but in \cite{fo33} 
the relativistic quantised field of the electron is never defined.
This could lead one to wonder whether \cite{fo33} is compatible with the definition used by Fermi and referenced in eq.~\ref{cicciolo}:
the answer is yes.\footnote{Assume that $\bm\psi=\sum_s \bm a_s \psi_s$ and exploit the formal symmetry of the theory.
For positive energies 
the state can be constructed 
from the vacuum  $|0\rangle$
by 
$|1_-\rangle= {\bf a^\dagger_{{ -}}} |0\rangle$;
for  negative energies 
one sets $|1_+\rangle= {\bf a_{{ +}}}\ |0\rangle$ instead;
the $\pm$ signs  remind us the charge of the particle. 
In this way, one is lead to define 
$
{\bm N}_{{ -}}=\bm a_{{ -}}^\dagger\ \bm a_{{ -}}$
and
$ {\bm N}_{{ +}}= \bm a_{{ +}} \ \bm a_{{ +}}^\dagger
$. 
Dirac's Hamiltonian will contain for each state a term
$ E (\bm a_{{ -}}^\dagger \bm a_{{ -}}  -  \bm a_{{ +}}^\dagger \bm a_{{ +}}) = E ({\bm N}_{{ -}}+  {\bm N}_{{ +}}-1)$, where $E=\sqrt{p^2+m^2}$. 
Compare also with Sect.~\ref{s415}.}
Crucially, the absence of an explicit definition for the full quantum field operator in their work highlights that their focus remained purely on reinterpreting existing particle states within the Hole Theory paradigm, rather than formulating a new, fundamental field theory.


\subsection{Heisenberg 1934 \cite{h34}}\label{ss:h}

Heisenberg's paper appears on May 21, 1934. It is entitled  
\begin{quote}
{\em Remarks on the Dirac theory of positron,} 
\end{quote}
 and cites both previous papers and a paper by Dirac \cite{d34}.\footnote{Dirac, in addition to claiming a prediction of positrons, presents his method of quantisation - that is, the Hole theory - highlighting the symmetry between electrons and positrons.}
In the second part, titled {\em Quantum theory of wave fields,} 
Heisenberg defines the quantised matter field as follows:
\begin{equation}\bm\psi(x,k)=\sum_{\mbox{\tiny for }E_n>0} \bm a_n\, u_n(x,k)    +\! \sum_{\mbox{\tiny for }E_n<0} (\bm a_n')^\dagger \, u_n(x,k)
\label{leben}
\end{equation}
where $k=1,2,3,4$ are the spin indices and $x$ are the space-time indices.\footnote{Eq.~\ref{leben} does not appear 
explicitly in \cite{h34}, but 
follows  replacing   in his eq.~48 the position given in 
eq.~53. Note that Heisenberg uses the notation  ($^*$) for the Hermitian conjugate, that we indicate with  
 ($^\dagger$) to  conform to modern notations.}   This differs somewhat from Fock's position,
and this time is consistent with Lorentz and gauge invariance, 
but still leads to the identification between operators of number $\bm N'_n=\bm 1-\bm N_n$,
just as suggested by Fock, Oppenheimer, and Furry (with a different but equivalent notation).
Heisenberg uses this field only to derive a density matrix operator, and he 
thanks Pauli for the remark that:  
{\em ``This representation of the density matrix agrees with the representation that was chosen
by Pauli and Peierls, Oppenheimer and Furry, Fock''.}\footnote{Pauli and Peierls' work was not published, but its content is described in \cite{papul}. Note that in his writings preceding  Majorana 1937,  Pauli admits  that he does not know how to treat fermions,  see also the subsequent Sections~\ref{ss:p} and Section~\ref{s:31}. It should be noted that in his 1941 review, Pauli will attribute these results entirely to Heisenberg.} In addition to the formal developments, this paper 
paves the way for subsequent calculations and results, in particular the work with Euler on the nonlinear interaction of light and light \cite{e36}.
Among the three papers discussed in  this section, this  is the one that will have the most impact.


\subsection{Majorana (1934?) \cite{supers}}\label{ss:dd}
Esposito's fascinating study of an unpublished French manuscript by Majorana \cite{supers}  revealed a report prepared for a seminar in the USSR. This report covered notable aspects of the ongoing scientific discussion of 1933-1934, covered in this section.   Thus, this manuscript chronicles the period during which Majorana withdrew from both Fermi's group and the outside world, publishing nothing for four years. This could be termed his `first disappearance', from which he decided to re-emerge in 1937,
to participate in a competition for a professorship.
It is highly plausible that these results depict some of the preparatory work for  \cite{m37};
moreover, the fact that this important work is not considered satisfactory, 
and therefore not published, is characteristic of Majorana's style.

\subsection{Pauli 1935  \cite{pp35}}\label{ss:p}
 Pauli presents the results of Heisenberg in a lecture at IAS with a very revealing title 
 \cite{pp35}:
\begin{quote}
{\em The Theory of Holes} 
\end{quote}
However, he adopts more vivid and eloquent symbols than Heisenberg, namely
 \begin{equation}\bm a \to \bm a^-\quad \mbox{and}\quad \bm a' \to \bm a^+\end{equation} 
In this manner, he emphasises that one refers to electrons, the other to positrons.
Quite importantly, he remarks that 
\begin{quote}
\small\em  ``We shall now consider Heisenberg's extension of Dirac's theory of the
positron. This theory is an attempt to find a new interpretation for the negative energy states, and makes an essential use of the exclusion principle.''
\end{quote}
A statement from the last sentence needs to be stressed:
\begin{quote}
\sf In Heisenberg 1934 \cite{h34} the fermionic character of the electron is not 
explained but assumed
\end{quote}
This is the key consideration that prompts Pauli and Majorana to pursue  the discussion, which will culminate in the developments examined in the next section. 

While the ``Dirac–Heisenberg'' procedure was formally sound and provided a consistent way to handle the mathematics of positrons, it remained ontologically rooted in the physical framework of the infinite Dirac sea. For the authors of this period, the modifications of the field were aimed at improving the features of the ensuing Dirac Hamiltonian, not at eliminating the hypothesis of the Dirac sea. The possibility of a theory that could be formulated without any reference to negative energy states — even as a background — remained outside the conceptual horizon of the time, only to be explicitly addressed in 1937.

%
%
%
%
%


\section{New procedures of quantisation. Conceptual clarifications (1934-1937)}\label{s:4}

During this period, systematic procedures were developed that represented a radical departure from Dirac's approach and demonstrated the possibility of quantising fields without relying on Dirac sea/Hole theory.
Pauli and Weisskopf, working from first principles, offer an alternative theoretical argument for the existence of anti-particles 
(Sect.~\ref{s:31}).
Majorana demonstrated that the quantum field of the electron 
must be fermionic  (Sect.~\ref{mm}), or recasting this in the form of an implication,
\begin{center}
\sc\small if the quantised field obey Dirac's equation, they  must \\
anti-commute to warrant the consistency of the quantum theory 
\end{center}
Majorana, in addition to proposing the new argument and a more coherent interpretation of the meaning of the fermionic quantum field, 
 confirmed Heisenberg's previously described position for the charged fermion field 
 and discovered a new possibility for neutral particles \cite{m37}. 
These findings will be usefully commented and expanded upon by other authors (Sects.~\ref{s:33} and~\ref{s:34}).
The period between 1934 and 1937 saw the Hole Theory become obsolete, marking a significant turning point in quantum field theory.

%
%

\subsection{Pauli-Weisskopf 1934 \cite{pw34}, Pauli 1936 \cite{p36}}\label{s:31}

Pauli and Weisskopf show how it is possible to formally construct 
a theory of a relativistic particle without spin, which at the time had no observational counterpart.\footnote{The application to Yukawa's theory
will be discussed in a paper issued on Sept.~25, 1937 \cite{ys}.}
It is not based on the exclusion principle, unlike the theory of holes, and yet it necessarily predicts the existence of
quanta of opposite charge just like Dirac's theory.
They motivate their investigation as follows
\begin{quote} \small \em 
``The main interest in the latter theory, it seems to us, lies in the fact that automatically
- without a new hole-like hypothesis and without any changes in the quantum theory
or strange cancelling limits and subtraction artifices$^{1)}$ - the energy of the particles is
always positive as a result of the quantisation of the fields.''
\end{quote}
It should be remarked that, in the parenthesis that explains the word ``automatically'', 
Pauli and Weisskopf distance themselves from several procedures considered {\em ad hoc} and unconvincing, 
including the one adopted in Heisenberg's paper. 
In fact, their  citation  is 
\begin{quote} \footnotesize \em 
$^{1)}$P.A.M. Dirac, Proc. Cambr. Phil. Soc. 30, Pt.II, 150 (1934).  --  R. Peierls, Proc.
Roy. Soc. A146, 420 (1934).  --  W. Heisenberg, Zeits. f. Phys. 90, 209 (1934).
\end{quote}
In plain words, the two authors declare themselves not entirely convinced by the developments of Dirac's theory described in the previous section.
Pauli will refer to this result of his as the ``anti-Dirac theory'' \cite{p35}. 

It is also interesting to read a passage of a letter 
that Pauli addresses to Heisenberg  
to present him this new result\cite{ssss}
\begin{quote} \small \em 
``The question of whether this theory can also be used to formally quantise matter waves 
using the exclusion principle required a more detailed investigation. Interestingly, the result is negative.''
\end{quote}
In 1936, Pauli will discuss this question in \cite{p36}, arguing that a scalar particle cannot obey
Fermi-Dirac statistics (i.e.,~cannot obey the exclusion principle).

 However, neither Pauli nor Weisskopf attempted to formulate a new theory for particles with half-integer spin. 
As we will see in his 1941 summary paper, Pauli simply describes the Dirac–Heisen\-berg procedure, thereby implicitly indicating that it is the best option of which he is aware.



A famous testimony by one of the Via Panisperna boys,
Giancarlo Wick, repeated over the years, indicates that Majorana had studied the quantisation of the scalar field at least three years before Pauli and Weisskopf, but without considering this result worthy of publication. This episode has been examined by Salvatore Esposito, who tested and consolidated its credibility \cite{salvus}.

\subsection{Majorana 1937 \cite{m37}}\label{mm}
Published in April, Majorana's paper sets out a theory of electrons and positrons which is consistent with both quantum physics and special relativity.  
The consistent application of quantum principles renders the Dirac Hole theory obsolete in both form and concept; however, Dirac's equation remains foundational.

To follow Majorana's argument, the modern reader must momentarily set aside the standard path of canonical quantisation. Majorana’s derivation is a masterclass in `reverse engineering': he identifies the only algebraic structure (anti-commutation) that allows a Hermitian field to satisfy the Dirac equation without vanishing energy.
Furthermore, I believe it is appropriate to immediately dispel any doubts modern readers may have about the purpose of Majorana's article. This is evident from the title, which translates as:
\begin{quote}\em 
 Teoria simmetrica dell'elettrone e del positrone
\end{quote}
which translates as:
\begin{quote}\em 
Symmetric Theory of the Electron and the Positron
\end{quote}
Therefore, the theory of neutrinos, for which many remember that work, is only one application.\footnote{In the title and elsewhere in the article, Majorana avoids using the term `anti-electron', which, {\em at the time,} referred only to the theoretical construct of hole theory. However, thanks to the development of fermionic quantum field theory, the term now has a different meaning and can be used without causing confusion, since hole theory has largely been forgotten (except, of course, in historical accounts).}

But let us proceed in order. 
The abstract efficiently summarises the paper's objectives and results
\begin{quote}\small\em
``It is shown how to achieve a full
formal symmetrisation of the quantum theory
of the electron and positron by making use of a new
quantisation process. 

The meaning of the equations of DIRAC
equations is quite modified and there is no longer any need to speak of states of negative energy.''
\end{quote}
Adopting evocative terms, we could say that it was Majorana who showed the world how to empty the Dirac sea.\footnote{It should be repeated that the Dirac sea hypothesis, unattractive from a physical point of view and now abandoned, is accompanied by relatively simple and much more spontaneous expressions for the second quantisation fields, Eq.~\ref{cicciolo}, where a wave and a quantum field are in close correspondence. 
This is why the Dirac sea hypothesis maintains a certain interest for didactic purposes:
it allows us to appreciate how we arrived at modern quantised field theory.} This achievement is precisely what gives the modern label of `symmetric quantisation’ its full meaning, as the theory treats particles and antiparticles symmetrically from the foundational level of the field equation, not merely as an ad hoc reinterpretation of existing states. Following this step, it becomes fully legitimate to speak of a ``vacuum state” rather than a ``ground state” as the Dirac sea.

Majorana begins the discussion from the Dirac-Heisenberg symmetric theory  --  he immediately quotes \cite{d34} and \cite{h34}  --   but remarks 
that their procedure is not entirely satisfactory 
\begin{quote}\small\em
``because they invariably start from an asymmetrical approach [...]
 Therefore, we have attempted a new path that leads more directly to the destination.''
 \end{quote}
 Majorana hypothesises that Dirac's equation for the electron must describe the behaviour of an operator  --  i.e., a quantum field  --  rather than that of a numerical function, and proceeds to demonstrate that it must obey anti-commutation relations, a property that characterises fermions.
He declares his thesis as follows, 
\begin{quote}\em\small
``The [field] equations can be made dependent on the Hamiltonian in the usual way,  
provided that appropriate anti-commutation relations are established between the quantum coordinates.''
\end{quote}
With the clause {\em ``in the usual way'',}  he means that Heisenberg's equation of motion should be obeyed to 
agree with the -- so to say, classical -- Dirac's wave equation.\footnote{The Hamiltonian is obtained by  a Hamilton density function that corresponds to the Dirac equation, and which in turn descends from a Lagrangian principle. However, following  Majorana, we should recognise the formal character of these manipulations, as they concern an operator - the field - and not  an ordinary function - the wave.}
After a very short and elegant discussion he displays the conclusion  in his eq.~(12): 
\begin{equation}\label{cordo}
 \bm U_i(q)\, \bm U_k(q')+ \bm U_k(q')\, \bm U_i(q) =  \delta_{ik}\ \delta(q-q')
 \end{equation}
which is the statement, in the formal language of field theory, that the 
quanta described by Dirac equation must 
 obey fermion statistics. As an evidence of the clarity of Majorana's thought, I also report his comment on the results that he obtains
 \begin{quote}\em\small
 ``These formulas are completely analogous, except for different statistics, to those obtained from the quantisation of Maxwell's equations.''
\end{quote} 

In \cite{m37} Majorana adopts a very transparent discretised  formalism for 
the 3-dimensional space and spinor indices, after having stated, 
\begin{quote}\small\em
``Leaving to the reader the obvious extension of the following formulas to the continuous systems of which we shall have to deal later, we set forth for greater clarity the quantisation method with reference to discrete systems.''
\end{quote}
%
 For the reader's convenience, the core of the formal argument for fermionic statistics is described below, making explicit the {\em ``obvious extension''} to the continuous system and adopting modern notation as in \cite{gdf}, a didactic paper; we have however replaced the symbol  $\Delta$  with
 $\hat{H}$  to conform to the notation used in \cite{lando} for the Dirac differential operator, given in eq.~\ref{dira}.


\paragraph{From Dirac equation to anti-commutation of Majorana fermions}


 
In Ref.~\cite{m37}  a suitable choice of basis for gamma matrices, $\gamma_\mu^*=-\gamma_\mu$ recalled in Sect.~\ref{2:2}, 
is used to simplify the discussion, after warning 
that for other choices of matrices some adjustments are needed.
At this point it is observed that the Dirac equation for a free particle (eq.~\ref{dira}),  that can be rewritten as  
$\partial \psi /\partial t=\hat{H} \psi/(i\hbar)$, 
allows us to treat the case of a real wave function, conveniently expressed thanks to Majorana's choice of gamma matrices by   $\psi^*=\psi$.
However, the ``classical energy density'' corresponding to the Dirac equation,\footnote{
Although this terminology appeared in later treatments (e.g., in \S 25 of \cite{lando}), Majorana introduces it in a critical context to highlight how a purely numerical (c-number) interpretation of the Dirac field leads to physically unacceptable results. 
The density in eq.~\ref{777} is corresponding quantum expression in Heisenberg representation.
The argument for its covariance is identical for classical real or complex wave functions and for 
fields satisfying the hermiticity condition given in eq.~\ref{ermione}.  
We introduce the coefficient $\nicefrac{1}{2}$ only to conform to modern notations, following \cite{gdf}; if we want, we can absorb it into the function $\psi$, as Majorana does. For maximum clarity, we show explicitly the spinor indices of the wave functions and fields $a,b,c...=1,2,3.4$.}
justified  from the invariant Lagrangian density  and that 
resembles that constructed for the electromagnetic field, namely
\begin{equation}\label{corresp}
\frac{1}{2} \psi^t \hat{H} \psi=
\frac{1}{2} \psi_a \hat{H}_{ab} \psi_b 
\end{equation}
gives  zero energy: this is pointed out after eq.~(3) of \cite{m37}.\footnote{Majorana \cite{m37} (pp.~172-173) rejects the c-number approach 
 as {\em poco soddisfacenti} (``somewhat unsatisfactory''). He proposes a  
{\em generalizzazione} (``generalization'') of the variational principle, 
where the field variables possess their {\em significato finale} (``final meaning'') as non-commuting operators from the outset. The expression for the Hamiltonan in his Eq.~(13) (p.~175),  the Hamiltonian of the new operatorial framework, follows from the variational principle of his Eq.~(11).}
The verification is straightforward:
The term in $\hat{H}$ proportional to the mass contains the antisymmetric matrix $\beta$.
The other term reduces to a total derivative, since 
$ 2 \psi_a \vec{\alpha}_{ab} (\vec{\nabla} \psi_b) = \vec{\nabla}(\psi \vec{\alpha} \psi ) $; assuming that the wave functions of the states have
negligible boundary values,
we arrive at the anticipated conclusion.\footnote{A consistency check is as follows: the 
Hamiltonian density must be real, but  it is $i\times$real; thus it is zero.
Interestingly Pauli repeats Majorana's observation four years later~\cite{p41} for complex wave-functions, see comment after his eq.~(81), but without recognising credit to his colleague. Although inserted in a different theoretical context, the mathematical coincidence of Pauli's observation underscores the persistence of the formal problem originally raised by Majorana.}
But at this point one may legitimately wonder, why should it not be possible to describe a real wave? 

(We can also say that this observation forces Majorana to move from a wave function to an operator. If, for the sake of argument, we felt compelled to use numerical functions to formulate it, the theory would run aground at Dirac's point. But precisely because we choose to formulate it with operators, it is precisely the ``magic” of the order of the terms — the lack of commutativity — that saves physics from the undesirable conclusion that follow from Dirac approach.)

 Majorana proceeds  declaring that we are not dealing with a wave function  --  a classical object  --  but rather with an operator-valued field  --  a quantum object. 
 (With our typographical convention, we will write: $\psi \to {\bm \psi}$.)
 Thus, the quantities described by the equations do not need to commute.
This simple  remark, consistently implemented in the light of quantum theory, allows a radical departure from previous theory;
it will turn out that, unlike the six Maxwell equations,  the four Dirac equations {\em do not} have a classical counterpart. 
Of course  the condition of reality is replaced with that of hermiticity:
\begin{equation}\label{ermione}
\bm\psi_a=\bm\psi^\dagger_a
\end{equation}
that defines the case of Majorana fermions.
Assume that the Hamiltonian is the operator
of the Hermitian field ${\bm \psi}$ that corresponds to eq.~\ref{corresp}:
\begin{equation}\label{777}
 {\bm H}=  \frac{1}{2}  \int {\bm \psi}_a(x) \  (\hat{H} {\bm \psi}(x) )_a\ d^3 x 
 \end{equation}
As a warm-up manipulation note that, integrating by parts  and discarding the total derivative term,  
still allows us to move the differential operator $\hat{H}$ to the left field,
$  \int {\bm \psi}_a(x) \  (\hat{H} {\bm \psi}(x) )_a\ d^3 x = 
 -  \int ({ \hat{H} \bm\psi(x)})_a \   {\bm \psi}(x) _a\ d^3 x$
(in the discretised presentation 
 adopted 
by Majorana,  this  amounts  simply to stating that 
the operator $\hat{H}$  is antisymmetric.)
 
 Unlike modern textbook approches that start from canonical quantisation, Majorana’s logical flow is inverted: he demands that the Dirac equation functions as the Heisenberg equation of motion for a quantum operator ${\bm \psi}$, and shows that this requirement uniquely dictates anti-commutation. Here is the key assumption,
 \begin{equation}\label{assunta}
i\hbar \frac{d}{dt} {\bm \psi}_b(y)= \hat{H}_{bc}  {\bm \psi}_c(y)
\stackrel{\mbox{\tiny Hp.}}{=}[ {\bm \psi}_b(y)\,  , \, {\bm H}]
\end{equation}
where, of course, on the r.h.s.~we have a commutator 
$[\bm{x},\bm{y}]=\bm{x}\, \bm{y}- \bm{y}\, \bm{x}$.
In plain words, the  electron field  --  a quantum object  --  should obey the equations that follow from quantum principles
(this corresponds to the discussion at page 174 of \cite{m37}).

 

Before we proceed, we would like to present a common-sense, algebraic argument as to why it is not too surprising that Majorana arrives at the anticommutation condition. This is precisely what is needed to transform Heisenberg's equation, cubic in the fields,  into Dirac's equation, which is linear instead: in fact, the left-hand side of equation~\ref{cordo} is bilinear in fields, whereas the right-hand side is a pure number.
Let us now examine the manipulations in detail.

\bigskip

Since the Hamiltonian is time independent,
we can assume that all fields are computed
at the same time $t$.
On the r.h.s.\ expression, we add and subtract the following term,
which can be rewritten by moving $\hat{H}$ to the left field (after discarding a total derivative) since this differential operator acts only on the integration coordinates $x$ and not on the~$y$:
\begin{equation}
\frac{1}{2}  \int {\bm \psi}_a(x)\  {\bm \psi}_b(y)\  (\hat{H}{\bm \psi}(x))_a \ d^3 x = 
-\frac{1}{2}  \int (\hat{H} {\bm \psi}(x) )_{a}\ {\bm \psi}_b(y)\  {\bm \psi}_a(x)\ d^3 x 
\end{equation}
This  does not introduce anything new, but it is helpful to emphasise the role of anti-commutators, as it will be evident just below.
In fact,  Majorana's assumption of eq.~\ref{assunta}
is shown to be equivalent to the following one\footnote{In \cite{m37} these formal steps are given after eq.~(5).}  
 \begin{equation}
i\hbar \frac{d}{dt} {\bm \psi}_b(y)=   \frac{1}{2} \int \Big(  C_{ab}(x,y) \ (\hat{H} {\bm \psi}(x)  )_a
+  (\hat{H} {\bm \psi}(x) )_a  \  C_{ab}(x,y)  \Big) d^3 x \label{20}
\end{equation}
where 
\begin{equation}
C_{ab}(x,y)= \{ {\bm \psi}_b(y) , {\bm \psi}_a(x)  \} 
\end{equation}
namely an anti-commutator $\{\bm{x},\bm{y}\}=\bm{x}\, \bm{y}+ \bm{y}\, \bm{x}$.
Evidently, the request of consistency can be  satisfied by setting in Eq.~\ref{20}
\begin{equation}
\{ {\bm \psi}_b(y) , {\bm \psi}_a(x)  \} = \delta^3(\vec{x}-\vec{y})\, \delta_{ba}
\end{equation}
(compare with eqs.~$(6),\, (6'),\, (7)$  and (12) -- shown above -- of \cite{m37}).
This argument indicates that the Hermitian field 
governed by the Dirac equation  must be anti-commuting, a
property that, in the formalism of Jordan, Klein, Wigner, Heisenberg and Fock, characterises fermionic particles.
%

\paragraph{The case of charged fields}
At this point, the quantisation of charged fields is obvious, but we provide the details here for the sake of completeness, still 
following the notations of \cite{gdf}.
Let us consider \cite{m37} the following construction
\begin{equation}\label{compos}
{\bm \psi}=\frac{{\bm \lambda}+i\, {\bm \chi}}{\sqrt{2}} \label{qce}
\end{equation}
where $\bm \lambda$ and $\bm \chi$ are the Hermitian quantised fields already defined, describing particles with the same mass.
Their anti-commutation relations imply
\begin{equation}
\{ {\bm \psi}_b^\dagger(y) , {\bm \psi}_a(x) \} = \delta^3(\vec{x}-\vec{y})\ \delta_{ba}
\end{equation}
The overall quantised field can be written as
\begin{equation} \label{piccin}
{\bm \psi}=\sum_s \left( 
{\bm c}_s \, \psi_{s} + 
\bar{\bm c}_s^\dagger \, \psi_{s}^* \right) 
\end{equation} 
with ${\bm c}_s\neq \bar{\bm c}_s$, where 
\begin{equation}
{\bm c}_s= \frac{{\bm a}_s + i {\bm b}_s }{\sqrt{2}}
\quad \mbox{ and } \quad
\bar{\bm c}_s= \frac{ {\bm a}_s - i {\bm b}_s }{\sqrt{2}} 
\end{equation}
compare with eq.~\ref{cicciolo}.
The free Hamiltonian density of ${\bm \psi}$ coincides with two independent free Hamiltonians densities 
\begin{equation}
\int {\bm \psi}^\dagger(x) \, \hat{H} {\bm \psi}(x) \ d^3 x
= \frac{1}{2}\int {\bm \lambda}^t(x) \, \hat{H} {\bm \lambda}(x) \ d^3 x +
\frac{1}{2}\int {\bm \chi}^t(x) \, \hat{H} {\bm \chi}(x) \ d^3 x
\end{equation}
The inclusion of electromagnetic interactions, obtained by replacing $\hat{H} \to \hat{H}_q$, leads to mixed terms
between these two fields, thus, we need both of them to represent a charged particle.\footnote{With the introduction of the generalised momentum of Fock 
\cite{jo}, even with the representation of Dirac gamma matrices by Majorana, 
we have the complex differential operator 
\begin{equation}
E\, \psi=\hat{H}_q \psi\mbox{ where }
\hat{H}_q= \left(c\vec{p}- q \vec{A} \right) \vec{\alpha}+ m c^2 \beta + q \varphi
\end{equation}
Note that
\begin{equation}\label{pazz}
(\hat{H}_{+q})^* = - \hat{H}_{-q}
\end{equation}
because the electromagnetic four-way potential $A=(\varphi,\vec{A})$ is
made up of real functions and the electric charge $q$ is also real, while both $E=i\hbar \partial/\partial t$ and $\vec{p}=-i\hbar\vec{\nabla}$ are purely imaginary. For more on this, see~\cite{wim}.}
We arrive at the so-called
Dirac Hamiltonian, but written using the quantised field ${\bm \psi}$. 
It is immediate to verify gauge invariance, i.e., the symmetry under 
$\delta{\bm \psi}=i \epsilon {\bm \psi}\Leftrightarrow \delta{\bm \lambda}=- \epsilon {\bm \chi} \, ,\, \delta{\bm \chi}=+ \epsilon {\bm \lambda} $.

 \paragraph{Reactions to the last paper of Majorana}
I would like to conclude this examination of \cite{m37} with some annotations on this paper and on 
 the scientific environment in which Majorana operated.\\  
Fermi's appreciation of this result can be ascertained from his judgement for the chair competition,
held in the same year (from \cite{e09}, preface, page xiii, translated from Italian \cite{recami}):
\begin{quote}
\small\em
``[Majorana] devised a brilliant method for treating the positive and negative electron symmetrically, finally eliminating the need to resort to the extremely artificial and unsatisfactory hypothesis of an infinitely large electric charge spread throughout space, an issue that had been addressed in vain by many other scholars''  
\end{quote}
Based on this work, the examining commission requested an exceptional merit position for Majorana from the Ministry of Education, Bottai. 
This was granted, and Majorana was appointed 
to the Institute of Professor Carelli at the University of Naples.\\
It is plausible this work came as a complete surprise to Fermi. 
To be sure, Majorana --- who held the title of {\em libero docente}, equivalent to  
{\em privatdozent} --- had submitted a proposal to teach a course in Rome in 1936, 
with the following program \cite{guro2}
\begin{quote}
\em\small Relativistic theory of the electron. Quantization processes. Field quantities defined by the laws of commutativity or anticommutativity. Their kinematic equivalence with complexes of an indeterminate number of individuals with Bose-Einstein or Fermi statistics, respectively. Dynamic equivalence.
Quantization of Maxwell-Dirac equations. Study of relativistic invariance. The positive electron and charge symmetry. Various applications of the theory. Radiation and diffusion. Creation and destruction of opposite charges. Fast electron collisions.
\end{quote}
This impressive list shows that the ideas drafted in 1934 had already reached full maturity, and the content of the work he would publish in 1937 was now ready. However, the faculty council, which included Enrico Fermi along with Corbino, Rasetti, Lo Surdo, and others, {\em did not} grant him equivalence to ordinary courses \cite{guro2}.  This amounted to academic ostracism  and most likely indicates that the value of Majorana's approach was not immediately understood.\\
Majorana's correspondence with Giovannino Gentile, collected in \cite{recami}, provides additional information that, among the other things sheds light on the reactions of the scientific community \cite{ponte}. One particularly interesting element can be found in a letter dated 25 August 1937, in which Majorana informs us that {\em ``a Swiss person had asked me for an extract of \cite{m37}''}. It is plausible \cite{ponte} that this refers to Stueckelberg, who was among the first to cite~\cite{m37} (in \cite{s38}). 
 \paragraph{Remarks on Majorana}
For an interesting analysis into  Majorana's logic and heuristics, see \cite{dra-esp}.
His way of thinking and his writing style, particularly discernible from \cite{m37},  
are those of a first-class mathematician \cite{bon1,guro}.
This trait, shared by few other visionary scientists of the time, sheds light on   the nature of his science 
 \cite{bs,bs1} and it is reflected in his teaching~\cite{asa,guro2}.\\
 To conclude, I would like to share a few thoughts on Majorana as an individual and his relationship with the scientific environment of his time. It is widely recognised that he was a highly individualistic scientist; however, there are sociological and epistemological aspects to consider, particularly regarding his interactions with the Rome and Leipzig groups \cite{bs1,bon1,guro}.
Majorana’s disposition notably shifted during his stay in Leipzig; he appeared more intellectually open and socially integrated, a change largely attributed to the profound rapport he established with Heisenberg. This period stands in stark contrast to his return to Italy, marked by a sudden and prolonged withdrawal from the Roman group. His subsequent four-year isolation (1933–1937) suggests that the pragmatic atmosphere of Via Panisperna might have been less conducive to his speculative nature than the theoretical depth he encountered in Germany. Furthermore, while his publications are the product of an intensely personal journey — which is particularly true of his 1937 paper \cite{m37} — they also represent a refined, albeit dramatic, dialogue with the foundational challenges of his era, the scientific world, and history itself.


\subsection{Ivanenko-Sokolov 1937 \cite{is37}}\label{s:33}
The  paper of Ivanenko-Sokolov is  received by the journal on April 14.
The two authors 
compare the quantised fermion field from Dirac-Heisenberg 
with the field of Pauli and Weisskopf 
\cite{pw34} emphasising their similarity and discussing their differences. 
They observe that,  as claimed by Pauli-Weisskopf, 
 if one applies Bose-Einstein statistics to Klein-Gordon equation, energy is positive definite;
 if one applies Fermi-Dirac, it is not;
 conversely, if one applies Fermi-Dirac statistics to Dirac equation, energy is positive definite;
 if we apply Bose-Einstein, it is not. 
 They conclude, 
\begin{quote}
\small\em
``From this point of view, Bose (or Fermi) statistics appears to be the most natural for the scalar (or Dirac) relativistic equation.''
 \end{quote}
Curiously, the quantised fermion field is written in components,  adopting Dirac representation of gamma matrices. 
Moreover, Ivanenko and Sokolov  {\em do not} quote Fock \cite{f33} or Oppenheimer and Furry \cite{fo33}. Even Heisenberg \cite{h34} is cited
only in the last part of the paper, the one devoted to applications.

\subsection{Racah  1937 \cite{r37}}\label{s:34}

Racah's article of 15 July 1937 is a reaction  to Majorana's paper. 
The specific and significant critique, particularly concerning Majorana's hypothesis that the neutron might be described as  `real' (Hermitian) particle (using Majorana language), is recalled in Appendix~\ref{appa}. 
The document still refers to ``solutions to Dirac’s equation with positive and negative energy,'' which seems to testify to the persistence of the conceptual framework that precedes Majorana’s.

Even if Majorana's quantisation procedure is not 
commented, this paper is relevant to the discussion of fermionic quantum fields.
Racah focusses on the new  hypothesis on neutral particles, that is reformulated  
in a language compatible with the formalism adopted by Fermi. 
This is done by replacing in the Hamiltonian of weak interactions the quantised fields of neutrinos $\bm\varphi^*$
(constructed {\em \`a la} Fermi) with a hermitian quantised field $\bm\varphi^*+C \bm\varphi$; the matrix $C$, which is included 
to maintain Lorentz invariance,  is borrowed from Pauli's treatment of Dirac matrices \cite{p36-2}.  


The acknowledgments contain an interesting testimony:\footnote{Recall that Racah had visited W.~Pauli in Zurich immediately after graduation (1931-1932), establishing a lifelong scientific association with him, comparable only to the one with his mentor E.~Fermi.} 
%
{\em 
I would like to thank Prof.~W.~PAULI for interesting and fruitful discussions on the topics of this note}. 
So we can assume that Pauli became aware of Majorana's work very early on.

 
 \subsection{Kramers 1937 \cite{k37}}\label{s:35}
On November 27 of 1937, 
Kramers published a paper\cite{k37} where he introduces the charge conjugation matrix for a generic representation of $\gamma$ matrices.
Majorana's article is cited, mentioning both the particular representation adopted \cite{cpt}
and also the new quantisation procedure. In the passage where Kramers  states
\begin{quote}
\small\em ``we will in this article represent some results of the MAJORANA-calculus (and, thereby,
also of the DIRAC-HEISENBERG formulation of the hole theory, with which it is practically equivalent) 
in a general form''
\end{quote}
he acknowledges the existence of Majorana's procedure. However, by insisting on framing it within the `hole theory' framework, Kramers implicitly highlights the persistence of the older conceptual horizon, demonstrating how long it took for the community to fully grasp that Majorana’s procedure was not just an equivalent technical variant, but a profound conceptual shift.\\
Recall that Pauli and Kramers had known each other for a long time.\footnote{See Ref.~\cite{pupul}  for an interesting remark about the discussions of the early days of quantum mechanics.} 
Pauli held Kramers' papers in high regard, and the two scientists corresponded from 1935 to 1938: 
the letters are collected in~\cite{papul}.

 
 \section{Developments and systematisation (1939-1941)}\label{s:5}
After 1939, the main protagonist is Wolfgang Pauli.  Building on and extending previous work (including the research conducted by Fierz and de Wet in their theses, Sect.~\ref{s41} and Sect.~\ref{s415}) he will produce two extremely influential works. The first concerns his famous argument on the
connection between spin and statistics (Sec.~\ref{s42}), the second is a summary paper on quantum field theory (Sec.~\ref{s43}).
In the second paper, Pauli would not offer significant new contributions to the quantum field of fermions, but rather
provide the reader with a systematic discussion of some previous results, recast in the form of a canon.\\
For information on this period, and in particular on the development of the spin-statistics connection, we refer to the exhaustive work of Blum \cite{b14}; our only significant addition is to draw attention to Majorana’s earlier results and the fact that they are selectively omitted in Pauli’s review work.

\bigskip
Furthermore, it is worth noting that one significant physical implication of Majorana’s work was immediately evident to the community: in 1938, Wendell Furry commented the paper of Majorana \cite{fu38} 
and in 1939 he pointed out the the 
new  hypothesis on the neutrino  allows for a new channel of double beta decay \cite{fu39}, which—at the time—was expected to be much faster than the standard process and thus easily detectable. This established that Majorana’s formulation was not only a formal development but had testable  applications to nuclear physics, bringing his work to the general attention during the very years in which Pauli was preparing his definitive synthesis on the spin-statistics connection.

\subsection{Fierz \cite{fi39}, Pauli-Fierz 1939 \cite{p39}}\label{s41}
In his habilitation thesis, supervised by Pauli, Fierz tackled the problem of quantising particles with arbitrary spin. This work, based on the free-wave equations, led to the remarkable result:
\begin{quote}
\small\em ``It turns out that particles with integer spin always approximate Bose statistics, and particles with half-integer spin always approximate Fermi-Dirac statistics''
\end{quote}
The special case where the spin is $\nicefrac{1}{2}$ is also discussed; however Majorana's work is not cited.

In \cite{p39} Pauli and Fierz obtain the wave equations for arbitrary spin in the electromagnetic field.
This investigation is preliminary to the quantisation of these particles; the paper of Majorana \cite{m37} is cited but only for the decomposition of the 
wave-functions into real and imaginary parts.

\subsection{de Wet 1940 \cite{dw40}}\label{s415} 
In the article from January 1940 based on his doctoral thesis, with the  title
\begin{quote}
\em
On the Connection Between the Spin and Statistics of Elementary Particles
\end{quote}
the young mathematician J.S.~de Wet, student of Howard Robertson,  makes some notable points.
The main observations of this paper are that:\\ 
1)~the anti-commutation relations allow to promote a class of  classical Hamiltonian densities into quantum theories, namely 
to put in correspondence the classical equations of motion and  the quantum ones, i.e., Heisenberg's equations of motion;\\
2)~only  the cases when Hamiltonian (and the Lagrangian) is linear in the derivative allow to avoid the conclusion that 
the field is trivial, $\bm\psi=0$.\\
The implications for the case of the Dirac equation (which is not mentioned explicitly) are quite clear.
Interestingly, the argument and conclusions are very similar to those of Majorana, despite him not being cited.

While de Wet’s derivation mirrors Majorana’s internal logic—establishing a rigorous correspondence between classical and quantum equations of motion—his contribution will remain peripheral. This lack of impact can be attributed not only to his status as a doctoral student but also to the overwhelming authoritative presence of Pauli. His systematic synthesis effectively acted as an epistemic filter, absorbing technical advancements while stripping them of the foundational weight that Majorana and de Wet had originally envisioned.

In the final part of the paper, de Wet considers the positivity of the quantum Hamiltonian
following the method of \cite{f33}, \cite{fo33}, and exhibiting the relations 
\begin{equation}
\bm N_- =\bm a_-^* \bm a_- \mbox{ and } \bm N_+ =\bm a_+ \bm a_+^* 
\end{equation}
Following Sokolov and Ivanenko, he does not cite any of the authors of  Sect.~\ref{s:3}.
The inclusion of electromagnetism is not discussed, and the possibility of an Hermitian fermion field is missed.
Note as a curiosity that this paper 
begins  with a decidedly sharp statement on Pauli,
\begin{quote}
\small\em ``Pauli  \cite{p36} attempted to show 
Fermi-Dirac quantization was not admissible for
the scalar wave equation. His work was not
correct but subsequently Sokolov and Ivanenko \cite{is37}
showed that the Einstein-Bose quantization of
the scalar wave equation that had been given by
Pauli and Weisskopf \cite{pw34} could not be extended to
the Fermi-Dirac case.''
\end{quote}

\subsection{Pauli 1940 \cite{p40}}\label{s42}
In the very famous paper of August 19, 1940, with affiliations Zurich and
Princeton, Pauli shows that the wave functions of
particles having half-integer spin are subject to Fermi-Dirac statistics, so that their
energy is positive definite; those with integer spin must obey Bose-Einstein statistics
so that observables at space-like distances commute. This generalises the point made previously for spin $\nicefrac{1}{2}$ (i.e., matter) particles by 
Majorana and by de Wet  to all possible cases with half-integer spin \cite{b14}.

His argument does not rely on the form of Dirac's equation; in connection, it is interesting to note that Pauli raises doubts on 
the necessity of a first-order equation in momenta. In fact, the third footnote of section 4 reads
\begin{quote}
\em\small ``The author therefore considers as not conclusive the original argument of Dirac, according to which the field
equation must be of the first order.''
\end{quote}
This may surprise a modern reader, but that could be perhaps understood in light of his extremely cautious attitude.\footnote{Even so, the fact that the quantised field has the same dimensions as the wave function means that the bilinear propagation term in the Lagrangian density must be a first-order term in momentum.}
%
%
%
%

In any case,  Pauli will  
return to using the Dirac equation
in a paper published a year later and discussed just below \cite{p41}, 
as he had already done in previous explorations of the spin-statistics connection~\cite{pb39}.
In \cite{p40} Majorana is not cited. De Wet's results for spins greater than 1 are dismissed, and those for spin \nicefrac{1}{2} are not commented on.

%

\subsection{Pauli 1941 \cite{p41}}\label{s43}
In his highly influent review paper, 
titled
\begin{quote}
\em\small Relativistic Field Theories of Elementary Particles
\end{quote}
Pauli bases  the discussion of quantised fields with half-integer spin on the Dirac equation.
 In the chapter that is relevant for our discussion,  
 entitled {\em Dirac positron theory (spin \nicefrac{1}{2})} he includes  three paragraphs, 
whose content we are going to examine.
Before doing that, two remarks are in order:
\begin{itemize}
\item 
The RMP of Pauli covers only a subset of approximatively 40  papers from previous scientific literature. This is clearly a carefully considered choice: while it helps readers to focus their attention, it also makes any omissions more significant. In the section on fermionic quantum fields the authors mentioned are Pauli (five times), Racah, Kramers, Heisenberg (twice), Dirac and Majorana (once).
\item Throughout the paper,  Pauli  adopts Dirac terminology -- recalled in Sect.~\ref{2:2} -- in order to distinguish clearly waves (c numbers) 
from operators (q number).
\end{itemize}

\paragraph{Paragraph $(a)$}
The first paragraph, titled {\em The c number theory} corresponds to non-quantised or classical field theory based on the Dirac wave function (c=commuting numbers)
with a summary of the properties of gamma matrices, largely based on~\cite{p36-2}.
\paragraph{Paragraph $(b)$}
This paragraph, titled {\em Quantisation in accord with the exclusion principle} is inspired by Dirac's hole theory,
as can be seen from the introduction,
\begin{quote}
\small  \em 
``The
vacuum therefore is defined as that total state
in which all individual negative energy states
are occupied.''
\end{quote}
The exposition begins with the statement,
\begin{quote}
\small  \em 
``This formulation of the Dirac theory of holes
is not entirely symmetric with respect to the two
particles of opposite charge. We shall follow
below a formalism proposed by Heisenberg 
which expresses the same physical content in
more symmetric manner.''
\end{quote}
which corresponds evidently to the stage of discussion of Sect.~\ref{s:3}, the one of `symmetric quantisation'.
Then the definition of fermionic quantum fields  is presented in two unnumbered equations: 
\begin{itemize}
\item The one before equation 93 simply imposes the anti-commutation  invoking the authority of Jordan and Wigner.
\item The one after equation 93a, where  the following 
analysis  of the field into oscillators is offered 
\begin{equation}
{\bm u}_\rho({x},x_0)
=\frac{1}{\sqrt{V}} \sum_k \sum_{ r=1,2} \{ {\bm u}_+^r( {k})\,  a^r_\rho({k})\, e^{ i [{ \vec{k}\cdot \vec{x}}- k_0 x_0]} + 
{\bm u}_-^{r*}( {k})\,  b^r_\rho(k)\, e^{ i [-  \vec{k}\cdot \vec{x} + k_0 x_0]} \}
\end{equation}
\end{itemize}
As in the previous literature, the momenta $k$ are discretised; $r$ is the helicity; 
$V$ is the formal volume of quantisation; $a^r_\rho(k)$ and $b^r_\rho(k)$ are spinors; 
${\bm u}_\pm^r(k)$ are operators; 
$\hbar=c=1$.
The final result is a field that coincides with the modern ones, except for Pauli's notations 
and minor details.
A comparison between this formula and the one given in Eq.~\ref{piccin} shows that the new presentation places less emphasis on the conceptual core of symmetric quantization, at the expense of details (i.e., the structure of free wave equations).


\paragraph{Paragraph $(c)$}
In the third paragraph, titled
{\em Decomposition with respect to charge conjugate functions. The case of a non-electric particle} 
Pauli writes 
\begin{quote}\small \em 
``This theory was first developed by E.~Majorana \cite{m37} in which use is made of the
special representation of the Dirac matrices with $\alpha$ real and
$C=I$ which is mentioned above. For the general case see
G.~Racah \cite{r37} and H.A.~Kramers \cite{k37}.''
\end{quote}
A lot of emphasis is placed on the neutrino hypothesis.
Great importance is given to the charge conjugation matrix, that Kramers introduces  just after Majorana's work \cite{k37}.
Majorana's argument \cite{m37} for the necessity of the anti-commutation relations 
is not indicated.
An unaware reader might get the impression that Majorana developed only the theory of neutrinos, not a direct procedure for quantising the electron field and an argument explaining its fermionic character.

Note that Pauli insists -- as in Ref.~\cite{p39} -- 
 on presenting Majorana's technique as a tool for {\em decomposing} complex fermion fields. We might be tempted to follow Pauli and assume that the complex field can be considered more important, but only if we accept
 an application-based perspective. However (as discussed in Sect.~\ref{mm})  from a theoretical (or even educational) perspective, real fermions are simpler entities, that help us to better understand the nature of charged fields: a pair of real (Majorana) fields enable the {\em composition} (construction, definition) of a complex field as shown in equation~\ref{compos}.
 This framing resulted in a canonical path-dependency: by presenting Majorana’s technique merely as a tool for decomposition, Pauli’s work established a hierarchical preference for complex fields. Regardless of Pauli’s personal motivations, the sociological effect was to obscure the real (Majorana) field as a primitive entity, relegating a fundamental insight to the status of a secondary computational device.
 
%

\subsection{Gentile Jr.~(1941) \cite{guro3}}\label{gjr}
On 13 September 1941, Giovanni Gentile Jr.~submitted a draft entry on Majorana for the {\sf Enciclopedia Italiana (Treccani)}, following a request made on 5 April 1940 \cite{guro3}. Regarding the 1937 paper and its genesis, Gentile wrote:
\begin{quote}
\em\small [Majorana] isolated himself in silence, re-emerging in 1937 to present his symmetrical theory of the electron and positron, which eliminated a fundamental difficulty in Dirac's theory.
\end{quote}
This phrasing accurately identifies the conceptual core of Majorana's final work, namely, the transition to a formal representation where the ``hole theory" (the Dirac sea) is no longer required. Notably, Gentile Jr. frames this paper in terms of electrons and positrons, without mentioning neutrinos. Due to the outbreak of war, and perhaps also to the early departure of Gentile Jr.~(1942), this contribution remained unpublished for over eighty years; it was finally recovered from the archives and integrated into the digital edition of the {\sf Enciclopedia Italiana} on 11 March 2026.\footnote{I would like to thank the Historical Archives of the {\sf Enciclopedia Italiana} for providing me with the primary sources relating to the earlier entries on Majorana. ``The text of the main online entry on Majorana was written by Ginestra Amaldi (Giovene) for the {\sf Enciclopedia Minore} on 29 October 1940, while the entry in the {\sf Second Supplement} was written by Quirino Majorana, Ettore's uncle, on 1 February 1949.'' My comments: The first attribution and its date clarify why Majorana's articles from 1932 and 1937 were listed together in the previous online version, under the premise that they both sought to address the issue of negative energies in Dirac's theory.  The second attribution explains why Dirac’s theory is not referenced, as Quirino was a well-known opponent of relativity. Notably, none of these three authors devoted a single word to the application of \cite{m37} to neutrinos.}

\subsection{Pauli 1946 \cite{p46}}
The Nobel Prize lecture for 1945,\footnote{A curiosity: the banquet speech was delivered
by Ivar Waller \cite{speq}, who in the 1930s contributed actively in the scientific discussion we are interested in, see for example the aforementioned work~\cite{w30}.} 
given by Pauli on 13 December 1946 
 \cite{1945}, does not belong to the period under investigation. However, it is a highly impactful document that  gives an account of the history of fermionic quantum fields.
 
\medskip

There, Pauli retraces the history of the understanding of the exclusion principle.
The first part (pages 27-38) corresponds to the content of the paper published in Science \cite{p46-p},  
written in preparation of the lecture.
There, Pauli begins with an enunciation of the exclusion principle, which was aimed at providing a meaning to the distribution of atomic electrons (1924).  He recalls its significance to theory of indistinguishable electrons (Heisenberg, Fermi, Dirac 1926), that imposed antisymmetric wave functions on a multi-electrons state. Pauli remarks that, together with the ideas of Bose and Einstein for the photon, this allowed progress towards a general classification of all known particles; both protons (1927) and neutrons (1933) were shown to belong to the same class as the electron. 
However, Pauli emphasises that he continued to perceive these results as {\em ``disappointing",} as they did not stem from the principles of quantum mechanics.

\medskip
The most interesting part for the discussion of fermionic quantum field in \cite{p46} 
 is however the second one, beginning on page 39.
There, Pauli goes beyond the  discussion  in \cite{p46-p} 
to encompass subsequent developments, arguing that the second significant advance in understanding the exclusion principle came with the theory of quantised fields.
He emphasises that 1)~he and Weisskopf constructed a field theory of particles without spin from first principles that feature anti-particles; 2)~he showed that particles cannot obey the exclusion principle and must instead obey the statistics of photons (1936). These results revealed a new theory of antimatter that was different from Dirac's theory, which was essentially based on the exclusion principle and the  Dirac sea hypothesis. 
Although Pauli mentions the symmetric quantisation procedure, he does not consider it to be particularly novel. He stresses that the fermion field has no classical counterpart and feels that the story is not yet over. 
The only research papers quoted that appeared after 1933 are his own.

\bigskip
A remark made at page 32 deserves to be emphasised 
\begin{quote}
\small\em
``I was unable to
give a logical reason for the exclusion principle or to deduce it from more
general assumptions. I had always the feeling and I still have it today, that
this is a deficiency.''
\end{quote}
At first glance, such a statement might raise the suspicion that Pauli had overlooked or misconstrued the arguments of Majorana and de Wet. However, it is far more plausible that Pauli’s persistent reluctance to accept the Dirac equation—extending beyond a mere rejection of hole theory—as a definitive foundation led him to disregard an entire class of foundational arguments. The consequence was a long-standing interpretative gap in the tradition: Pauli’s authoritative `filtering' established a monopolistic narrative over the spin-statistics connection, effectively stalling a broader recognition of the internal logic already established in the Majorana-de Wet approach.

\begin{table}[t]
\centering
\caption{\small References to primary literature in influential secondary texts regarding the origin of spin \nicefrac{1}{2} quantisation. The top section lists textbooks; the bottom section lists historical accounts. The fourth column highlights the Majorana fermion quantisation procedure, typically omitted in these works. Note that references to Pauli: \cite{pw34,p36} (before Majorana 1937) and \cite{p40,p41} (after Majorana 1937).}\label{checz}.
\label{tab:secondary_lit}
\vskip3mm\small
\begin{tabular}{@{}llccc@{}}
\toprule
\textbf{Author} & \textbf{Date} & \textbf{Previous} & \textbf{Reference to} & \textbf{Subsequent} \\
\textbf{and reference} & & \textbf{references} & \textbf{Majorana \cite{m37}} & \textbf{references} \\ \midrule
\textit{Textbooks} & & & & \\
Heitler \cite{h36} & 1936 & \cite{jw,h34,p36} & -- & -- \\
Wentzel \cite{w43} & 1943 & \cite{jw,f33,fo33,h34,pw34} & no & no \\
Schweber \cite{s61} & 1961 & \cite{jw,pw34,p36} & no & \cite{p40} \\
Bjorken, Drell \cite{bd65} & 1965 & \cite{jw,p36} & no & \cite{p40} \\ \midrule
\textit{Historical accounts} & & & & \\
Brown, Hoddeson \cite{b83} & 1983 & \cite{jw,fo33,h34,pw34} & no & \cite{p41} \\
Pais \cite{p86} & 1986 & \cite{jw,f33,fo33,h34,pw34} & (neutrino, $\gamma^*_\mu=-\gamma_\mu$) & no \\
Schweber \cite{s94} & 1994 & \cite{jw,f33,fo33,h34,pw34} & (Schwinger's lecture) & \cite{p40,p41} \\
Jacob \textit{et al.} \cite{mj} & 1998 & \cite{fo33,h34,pw34} & yes & \cite{p40,p41} \\ \bottomrule
\end{tabular}
\end{table}

\section{Summary and final considerations}\label{s:6}

This paper has re-examined the pivotal transition in high-energy theoretical physics between 1933 and 1937, a period that laid the foundations for the modern concept of the fermionic quantum field. By systematically analysing the primary literature, we have reconstructed the shift from the early ``quantum waves of matter" approach to a structured canon of quantisation procedures. Central to this evolution was the gradual superseding of Dirac’s Hole Theory, which evolved into a coherent field theory through several intermediate stages, most notably the 
``Dirac-Heisenberg" phase.

A key finding of this study is the decisive, yet historically under-appreciated, role played by Ettore Majorana. While his 1937 paper is often cited for its specific applications, we have shown that it represented a reasoned and mature approach to the quantisation of half-integer spin particles. Majorana provided a compelling argument for why field operators must satisfy anti-commutation relations, effectively demonstrating that the Dirac equation could be interpreted without recourse to the ``Dirac sea."

Despite this conceptual leap, the recognition of Majorana’s contribution remains remarkably partial in the secondary literature, 
as it is clear from Table~\ref{checz}.
A glaring--and, literally, visual--example of such an absence is given by the historical synoptic table in the introduction of \cite{s94}.
Another striking example is found in Duck and Sudarshan’s authoritative book on the spin-statistics theorem \cite{suda}: While they rightly praise the quantisation procedure in de Wet’s article,\footnote{The interested reader can refer to  \cite{suda} for a discussion of why de Wet’s argument regarding the connection between spin and statistics does not extend directly from the case of spin $\nicefrac{1}{2}$ to the general case.} Majorana’s role is relegated to specific $\gamma$ matrices or the neutrino hypothesis. 

This highlights a persistent gap in the tradition: the failure to acknowledge that Majorana’s ``Symmetric Theory of the Electron and the Positron" was a general and fundamental breakthrough, not a specialised case for neutral fermions.\footnote{Note, for comparison, the  contributions of most authors cited above and appering in Table~\ref{tub1} --  including Jordan, Heisenberg, Weisskopf, in more recent times  
Fock, Oppenheimer, Furry, but mainly Pauli are recognised; 
e.g.,  Duck and Sudarshan devote an entire chapter to Pauli and Weisskopf results.}
The only  exception to this rule of which I am aware 
is the account of Maurice Jacob, a student of Francis Perrin and Gian Carlo Wick, 
who knew well Majorana and his story: see the last line of Table~\ref{checz}.

This historical underestimation likely stems from a combination of factors: the linguistic barrier of his original paper in Italian, the differing research priorities of the Fermi group, and his mysterious disappearance in 1938. Furthermore, the subsequent dominance of Pauli’s more complex (though ultimately equivalent) 1941 procedure served to obscure the clearer, more direct arguments presented by Majorana.

In conclusion, restoring Majorana’s work to its proper place is essential for a complete historical understanding of quantum field theory. By moving beyond the construct of ``holes," Majorana did not merely refine a technique; he established the framework for the modern fermion field, a contribution that deserves full integration into the historical narrative of 20th-century physics.

\paragraph*{Acknowledgments}
{\small
With the partial support of the research grant 2022E2J4RK {\em PANTHEON: Perspectives in Astroparticle and Neutrino THEory with Old and New Messengers,} PRIN 2022 program funded by the Ministero dell'Università e della Ricerca'' (MUR). Based on the seminars of January 16, 2025, at the University of Hamburg and on \cite{gdf,ponte,bang}. I thank Zurab Berezhiani, Alexander Blum, 
Salvatore Esposito, Roberto Lalli, Orlando Luongo, Guenter Sigl, Oleg Yurievich Smirnov, Andreas Ringwald, and Roberto Soldati for valuable discussions. I am particularly grateful to the two anonymous referees of EPJH for their constructive critical remarks and excellent advice.
}

\appendix

\section{The 1937 paper of Majorana: notes for a modern reader}\label{appa}
Ettore Majorana's 1937 work is a masterpiece of theoretical physics, but it is difficult to read. Modern readers should distinguish its revolutionary core from its real and perceived weaknesses. This appendix aims to provide easier access to the heart of the work.
 
 \subsection{General aspects}
 
\paragraph{Contextual and formal barriers} 
A primary difficulty lies in his presumption of high mathematical maturity of his readership. Majorana operates within the universally known ``Majorana Representation,” leaving ``appropriate modifications" for general representations as an implicit exercise. Here are his words on the point
\begin{quote}
\em\small ``We shall refer precisely to a system of intrinsic coordinates such as to make eq.~(8) [Dirac's equation] real, expressly warning that the formulas we arrive at are not valid, without appropriate modifications, in general coordinates.''
\end{quote}
Understanding this representation is a necessary first technical step to accessing the core of the work, but it is not
the key point of the work.
Some essential notes on this subject are given in Sect.~\ref{2:2}; 
for further introduction, see the subsection just below.
\newline
Another barrier is Majorana’s heavy emphasis on dismantling Dirac’s ``hole theory.” It is essential to view this not as a localized historical polemic, but as the final act of an intellectual battle begun five years earlier. Majorana's quantisation procedure aims to make negative energy states logically redundant, completing the conceptual construction of quantum field theory (QFT) that had already been initiated; note that this modern term was not yet in use.\\
Some barriers stemming from the 1937 professorship competition deadlines have only a formal character: 
E.g., the first reference is misdated: P.A.M. Dirac's article in \textit{Proc.~Camb.~Phil.~Soc.} is indicated as having been published in 1924, when the correct date is, obviously, 1934.

\paragraph{The odd hypothesis on the neutron} 
In proposing the identity between the neutron and the antineutron (the condition $\bm\psi = \bm\psi^{\mbox{\tiny C}}$ in modern formalism), Majorana demonstrated a striking detachment from the 1937 empirical landscape.
This approach, that prioritised the internal mathematical symmetry and formal elegance over the evidence from the data, 
most likely reflects Majorana's 1934–1937 intellectual seclusion, where he prioritized internal mathematical symmetry over the era's emerging phenomenological data.\newline
But this position is anomalous. Majorana, who had surgically corrected Heisenberg’s 1933 nuclear model, in 1937 seemingly ignored glaring experimental evidence contradicting his hypothesis:
\begin{itemize}
    \item \textit{Magnetic moments:} A self-conjugate particle implies a vanishing magnetic moment, contradicting non-zero values measured by the Stern group.
           \item \textit{Nuclear stability:} This identity is empirically untenable as it permits anomalous nuclear transformations that threaten atomic nuclei stability.\footnote{The $\beta^+$ decay $n \to \bar{p} + e^+ + \bar{\nu}$ would permit anomalous nuclear transformations such as $    (A, Z) \to (A - 2, Z-1) + e^+ + \bar{\nu}+\gamma$s, 
   implying violation of the Fajans-Soddy's law of $\beta$ decay
   and emission of GeV energy $\gamma$-rays.} 
\end{itemize}
 Whatever the reasons, Majorana paper deserved the criticism of  Racah:
the $n = \bar{n}$ hypothesis, in its simplest form, cannot apply to the neutron.
(Recall that Giulio Racah and Gian Carlo Wick participated in the same 1937 ``concorso'' as Majorana. Notably, Wick had recently 
provided a seminal discussion on the neutron's anomalous magnetic moment \cite{wick2}. This proximity makes Majorana’s disregard for these specific phenomenological constraints  even more striking.)

\paragraph{Summary}
In conclusion, putting aside the odd hypothesis on the neutron, 
the perceived `weaknesses' of the 1937 paper stem from an austere economy of style, not internal inconsistency. Majorana provided essential formal tools—most notably the construction of complex fermion fields from quantized real ones. Far from being a flawed exercise, the work remains an indispensable primary source for understanding the foundations of modern QFT.
As for phenomenology, it is precisely this work that sparked interest in Majorana fermions and, in particular, Majorana neutrinos, which are being researched in laboratories around the world.

 \subsection{Majorana representations of $\gamma$ matrices}
The form of the $4\times 4$ matrices chosen by Majorana helps us clearly understand the connection between the Dirac equation and the existence of anti-electrons. However, since this notation is not widely used, it is worth taking a moment to familiarise yourself with it. The following notes   on this topic, based on \cite{gdf}, supplement those in the appendices of~\cite{wim, oscio}.

\paragraph{Definitions and notation}
We begin by recalling the relationship between the (Dirac) $\gamma$ matrices and the $\alpha, \beta$ matrices:
\begin{equation}
\gamma^0 = \beta, \quad \vec{\gamma} = \beta \vec{\alpha} \quad \Longleftrightarrow \quad \beta = \gamma^0, \quad \vec{\alpha} = \gamma^0 \vec{\gamma}
\end{equation}
where $(\vec{\gamma})^i = \gamma^i$ ($i=1,2,3$). All representations of the gamma matrices are physically equivalent (Pauli's theorem~\cite{p36-2})\footnote{See  \cite{oscio} for other proofs.}, provided they satisfy the (Clifford) algebra:
\begin{equation}
\{\gamma_\mu, \gamma_\nu\} = 2 g_{\mu\nu} \mathbb{1}_{4\times 4}, \quad \mathrm{ with  }\quad g = \mathrm{diag}(+1,-1,-1,-1)
\label{gm}
\end{equation}
and the hermiticity conditions $(\gamma^0)^\dagger = \gamma^0$ and $(\gamma^i)^\dagger = -\gamma^i$.

\begin{table}[t!]
\centering
\caption{\small Three Majorana forms of the $\gamma$ matrices, all having the same
expression as $\gamma^3$ (which is diagonal), expressed as tensor products of Pauli matrices.
In  each of the forms shown here, we can
 $i$) change the sign of any matrix;  
 $ii$) swap the position of the last three matrices~$\gamma$  among themselves as desired;
 $iii$)~exchange the indices 1$\leftrightarrow$3 for all 4 matrices, or  the two Pauli matrices in the tensor product.}
\label{tab111}
\vskip3mm\small
\begin{tabular}{l cccc} 
\toprule
Form & $\gamma^0$ & $\gamma^1$ & $\gamma^2$ & $\gamma^3$ \\ 
\midrule
(a) & $\sigma_2 \otimes \sigma_1$ & $i \sigma_2 \otimes \sigma_2$ & $i \sigma_1 \otimes \sigma_0$ & $i \sigma_3 \otimes \sigma_0$ \\ \addlinespace
(b) & $\sigma_1 \otimes \sigma_2$ & $i \sigma_1 \otimes \sigma_1$ & $i \sigma_1 \otimes \sigma_3$ & $i \sigma_3 \otimes \sigma_0$ \\ \addlinespace
(c) & $\sigma_2 \otimes \sigma_0$ & $i \sigma_1 \otimes \sigma_1$ & $i \sigma_1 \otimes \sigma_3$ & $i \sigma_3 \otimes \sigma_0$ \\ 
\bottomrule
\end{tabular}
\end{table}

\paragraph{Spinorial realization of the Lorentz group}
Spinor wave-functions and fields $\psi$ transform under Lorentz transformations as 
\begin{equation}\psi' = \Lambda(\omega)\ \psi,
\end{equation}
 with the matrices:
\begin{equation}
\Lambda(\omega) = \exp\left( -\frac{i}{4} \omega_{\mu\nu} \Sigma^{\mu\nu} \right), \quad \Sigma^{\mu\nu} = \frac{i}{2} [\gamma^\mu, \gamma^\nu]\label{evidently}
\end{equation}
The generators $\Sigma_{0i}$ are anti-hermitian and generate boosts, while $\Sigma_{ij}$ are hermitian and generate rotations; $\omega$ are the six real (Lie group) parameters—rapidity parameters and (Euler) angles. 
For the (Dirac) adjoint 
$\bar{\psi} \equiv \psi^\dagger \gamma_0$, the transformation law is 
\begin{equation}
\bar{\psi}' = \bar{\psi}\ \Lambda(\omega)^{-1}=
\bar{\psi}\ \Lambda(-\omega).
\end{equation}

\paragraph{Explicit Majorana representations}
Table~\ref{tab111} provides several Dirac $\gamma$ matrices satisfying the (Majorana) condition:
\begin{equation}
{\gamma}_\mu^* = - {\gamma}_\mu, \quad \mu=0,1,2,3
\label{mjr}
\end{equation}
They are constructed via tensor (Kronecker) product ($\otimes$)  of  $2\times 2$ hermitian matrices, where  
 $\sigma_0=\mathbb{1}$ is the identity matrix and $\sigma_{1,2,3}$ Pauli matrices.
 It is easy to test the anticommutation rule and the hermiticity properties. 
 Note that the matrices $\Lambda(\omega)$ in Eq.~\ref{evidently} are real.

\section{Reports of SIPS from 1937-1939}\label{appb}

A poignant way to reconstruct the atmosphere in Italy during the publication of Majorana’s final paper \cite{m37} is to examine the Proceedings of the Italian Society for the Advancement of Science (SIPS). By the late 1930s, the SIPS had been largely co-opted by the Fascist regime to exert ideological and political control over the scientific community. The reports from the annual congresses in Venice (1937), Bologna (1938), and Pisa (1939)—all published in Italian—are characterised by nationalistic rhetoric and a heavy emphasis on military applications of research. In these accounts, scientists are often valued more for their perceived social utility to the state than for their intrinsic intellectual contributions.

While physics did not always take center stage in these proceedings, several key figures from Majorana’s circle remained prominent within the organisation. Enrico Fermi is listed as a SIPS vice president during this period. Antonio Carrelli, who would later host Majorana in Naples, presided over the physics section from 1937 to 1938 before being succeeded by Franco Rasetti (1939).
The general reports on physics for this period, authored by Giulio Dalla Noce, offer a snapshot of how Majorana’s work was initially received. Dalla Noce included an abstract of Majorana’s 1937 paper in his discussion of ``general theories," notably placing it on the same level as Giulio Racah’s commentary. 

The proceedings also reflect the shifting fortunes of the era: in 1937, Rasetti is seen requesting additional research resources (with little apparent success), and by 1938, Carrelli is found formally mourning Majorana’s mysterious disappearance. Notably, Fermi’s 1938 Nobel Prize is absent from these volumes, likely due to the political sensitivity following his flight to the United States. The 1939 report includes a lecture by Giovannino Gentile Jr.~(a close friend of Majorana and his only collaborator) on Dirac’s relativistic wave equations for particles with arbitrary spin, a topic that directly intersects with the theoretical concerns of this paper.
On the same occasion his father, Giovanni Gentile -- a powerful and influential figure in the fascist regime -- proposed and presided over the foundation of the {\em Domus Galileiana}, that today hosts  the collection {\em Fondo E.~Majorana}  --  the most important resource for accessing his unpublished studies.

\newpage
{\small

 
  \newpage
 { \sf \tableofcontents}

\end{document}